\documentclass[conference,compsoc]{IEEEtran}

\usepackage[T1]{fontenc}
\usepackage[utf8]{inputenc}
\usepackage{graphicx}
\usepackage{booktabs}
\usepackage{array}
\usepackage{amsmath,amssymb}
\usepackage{url}
\usepackage{balance}
\usepackage{placeins}
\usepackage{float}
\usepackage[hidelinks]{hyperref}
\usepackage{xcolor}

\newcommand{\hd}[1]{\textbf{#1}}

\begin{document}

\title{Recording-Aware Evaluation of\\Multimodal OT Intrusion Detection}

\author{\IEEEauthorblockN{Mustafa Umut Ozbek, Khalil El-Khatib, Li Yang}
\IEEEauthorblockA{Ontario Tech University, Oshawa, Ontario, Canada\\
Emails: mustafaumut.ozbek@ontariotechu.net,
khalil.el-khatib@ontariotechu.ca, li.yang@ontariotechu.ca}}

\maketitle

\begin{abstract}
Machine-learning-based intrusion detection systems (IDSs) for operational
technology (OT) are commonly evaluated on benchmark datasets in which normal
and attack data are collected in separate recordings. In such settings, high
performance may reflect recording-associated differences as well as attack
behavior. We introduce a recording-aware evaluation procedure that compares
attack windows with background from the same recording, uses prior data for
model fitting and calibration, tests temporal alignment against randomized event
schedules, and reports event coverage together with alarm burden. On the
ProvICS multimodal dataset, cyber modalities strongly distinguish the benign and
attack recordings before the first labeled attack. Same-recording evaluation
reverses the apparent modality ranking across all studied window sizes: the
physical view becomes the strongest individual modality, while cyber-only fusion
shows the largest difference between cross-recording and same-recording AUC,
reaching 0.54. Operational analysis further shows that permissive event-overlap
metrics can overstate detection when a small number of long alarm episodes cover
multiple actions. An external analysis on HAI shows smaller and reference-dependent
recording sensitivity. These findings show that cross-recording benchmark performance should be
interpreted alongside same-recording, forward, temporal, and operational
evaluations.
\end{abstract}


\section{Introduction}
\label{sec:intro}

Operational technology (OT) systems monitor and control physical processes in
critical infrastructure, where reliability, availability, and safety are central
requirements~\cite{nist80082r3,iaea33t}. Artificial intelligence and machine
learning (AI/ML) are increasingly used in OT intrusion detection systems (IDSs)
to identify suspicious behavior from network traffic, host and controller
activity, and physical-process measurements~\cite{umer2022}. These systems are
commonly evaluated on benchmark datasets using metrics such as AUC, precision,
recall, and F1.

High benchmark performance, however, does not necessarily reveal what
information a detector has learned. In some OT datasets, normal operation and
attack activity are collected in separate recordings. When those recordings
differ in operating state, configuration, traffic patterns, acquisition
conditions, or other contextual properties, the differences can become
correlated with attack labels. A detector may therefore separate normal and
attack samples partly by learning the recording boundary rather than
attack-specific behavior. This issue is especially relevant to multimodal IDSs,
where several data sources may share recording-associated differences and fusion
may reinforce them.

We study this problem through recording-aware evaluation. In addition to
conventional cross-recording evaluation, it compares attack windows with
background windows from the same recording. It uses only prior data for model
fitting and calibration, checks whether detector scores are temporally aligned
with labeled attack periods, and evaluates alarm burden and event detection
separately. Together, these checks clarify whether high benchmark performance
remains meaningful after accounting for recording and temporal structure. We
use recording to mean one continuous data capture.

The procedure is model-agnostic: it is intended to assess whether reported
AI/ML-based IDS performance reflects attack-specific evidence rather than the
recording structure of a benchmark.

This work complements research on experimental bias, temporal leakage, unseen
attacks, cross-network generalization, and dataset quality
\cite{arp2022,tesseract2019,kus2022,sok2023,engelen2021,silva2022,micem2025}.
Rather than proposing a new IDS architecture, we focus on the evaluation
procedure itself. ProvICS is the primary multimodal case study because it
provides synchronized host, controller, network, and physical-process evidence
together with ordered attack campaigns. HAI provides an external check of
whether similar recording-related differences appear in another OT benchmark.

We address three research questions: (RQ1) Can recording-associated
differences be detected before any labeled attack occurs? (RQ2) How does
same-recording evaluation change the apparent performance of individual
modalities and multimodal fusion? (RQ3) Do these observations remain
meaningful under forward, temporal, and operational evaluation?

This paper makes three contributions. First, it introduces a recording-aware
evaluation procedure that combines same-recording comparison, forward fitting
and calibration, temporal randomization, and event-level alarm evaluation
(Sec.~\ref{sec:protocol}). Second, using ProvICS, it shows that cyber modalities
strongly distinguish benign and attack recordings before the first labeled
attack and that same-recording evaluation reverses the apparent modality ranking
across multiple window sizes (Sec.~\ref{sec:results}). Third, it examines the
model-family sensitivity and operational meaning of these findings through supervised models,
alarm-level evaluation, temporal controls, and an external check on HAI.\footnote{Code and experiment configurations for this paper are available at: {\urlstyle{same}\url{https://github.com/ANTS-OntarioTechU/Recording-Aware-Evaluation-of-Multimodal-OT-Intrusion-Detection}}.}

\section{Related Work and Evaluation Gap}
\label{sec:related}

\subsection{Intrusion Detection in OT: From Rule-Based to AI/ML Methods}
OT intrusion detection has traditionally relied on signatures, rules, and
engineering knowledge to identify known unsafe or malicious behavior.
AI/ML-based methods complement these mechanisms by learning patterns from
historical system behavior and can surface deviations not captured by fixed
rules~\cite{nist80082r3,umer2022}. In OT, these methods use evidence from both
cyber and physical components of industrial systems. Process-based approaches model
sensor and actuator behavior using benchmarks such as SWaT, WADI, and
HAI~\cite{swat2016,wadi2017,hai2020}. More recent methods capture
relationships among process variables through graph-based and physics-guided
learning~\cite{pdgat2025,pcgat2025,physguided2026}. Provenance datasets and
traces provide host-focused evidence complementary to network and protocol
data~\cite{cicapt2024,darpatc}.

Multimodal approaches combine several of these evidence sources to improve
detection. Prior work includes correlation-based methods~\cite{jadidi2023},
attention-based autoencoders~\cite{mfgan2024}, and synchronized learning across
network and process data~\cite{kim2026}. These methods demonstrate the value of
combining heterogeneous OT information, while ICSSIM supports realistic
security-testbed development~\cite{icssim2023}.

ProvICS~\cite{provics2026} is particularly relevant because it provides
synchronized supervisory-host provenance, PLC provenance, decoded Modbus
traffic, and physical-process measurements on a common timeline. We use these
synchronized views to examine how benchmark recording structure affects the
interpretation of IDS performance.

\subsection{Evaluation Validity in Security ML}
Reliable evaluation of security-ML systems requires more than classification
metrics. Attack prevalence affects the operational meaning of IDS alarms
\cite{sommer2010,axelsson2000}, while experimental design can introduce temporal
leakage or other forms of bias. Arp et al.~\cite{arp2022} identify recurring
evaluation pitfalls, including spurious correlations, in machine-learning security research, and
TESSERACT~\cite{tesseract2019} shows how performance can change when temporal
and spatial constraints are enforced.

Related work has examined several forms of generalization in industrial
intrusion detection. Kus et al.~\cite{kus2022} study attacks excluded from
training, while FANCY-X~\cite{fancyx2026} evaluates generalization across OT
networks. Other studies highlight heterogeneity in IDS datasets and evaluation
procedures~\cite{sok2023}, identify dataset-generation artifacts
\cite{engelen2021}, examine intrinsic class separability~\cite{silva2022}, and
propose indicators for assessing dataset suitability~\cite{micem2025}.

These studies address complementary questions about time, attack type, network
environment, and dataset construction. We consider an additional aspect of
benchmark design: recording structure. When normal and attack data originate
from different recordings, a detector may exploit recording-associated
differences correlated with the labels. Recording structure therefore introduces
a distinct evaluation question: does the observed separation persist when attack
windows are compared with background from the same recording? In this
comparison, the fitted detector and attack-positive windows remain unchanged;
only the background source changes. The recording-membership probe is a
domain-classification check, often termed adversarial validation. The
contribution is the coordinated use of established checks to connect recording
separability with paired discrimination, temporal alignment, and alarm behavior
in multimodal OT data.

\section{Recording-Aware Evaluation}
\label{sec:protocol}

The purpose of recording-aware evaluation is to determine whether apparent IDS
performance depends on how normal and attack data are distributed across
recordings. The procedure does not replace conventional benchmark evaluation.
Instead, it adds complementary checks that separate cross-recording
discrimination from same-recording attack discrimination, preserve
chronological order during fitting and calibration, test whether detector scores
align with labeled attack periods, and evaluate alarms at the event level.

\begin{figure*}[t]
\centering
\includegraphics[width=\textwidth]{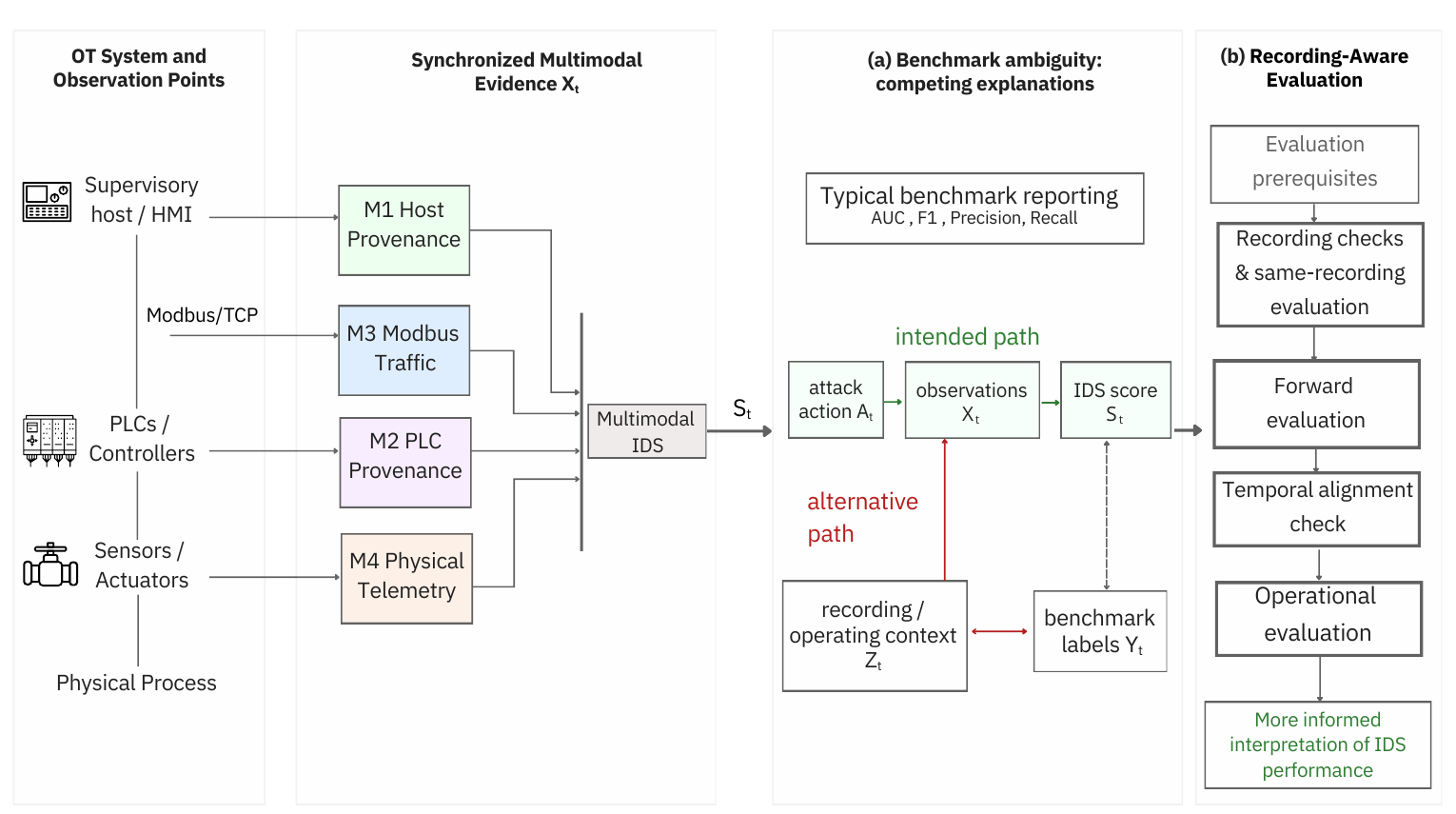}
\caption{Multimodal OT evidence and recording-aware evaluation. Host, controller,
network, and physical-process evidence are aligned as observations for an IDS.
When recording context co-varies with labels, cross-recording scores can reflect
either attack evidence or recording-associated variation. The workflow makes that
alternative testable through recording-aware, forward, temporal, and operational
checks.}
\label{fig:protocol}
\end{figure*}

\subsection{Why Cross-Recording Performance Can Be Misleading}
\label{sec:validity}
Consider a benchmark in which normal operation is collected in one recording
and attack activity in another. If the recordings differ in operating state,
system configuration, network behavior, acquisition conditions, or other
contextual properties, those differences can become correlated with attack
labels. A detector may therefore achieve high benchmark performance by
responding partly to recording-associated information rather than to
attack-specific behavior. Let $A_t$ denote a labeled attack action, $Y_t$ its
benchmark label,
$X_t=\{X_t^{(1)},\dots,X_t^{(M)}\}$ the multimodal observations, and
$S_t=f_\theta(X_{\le t})$ a continuous IDS model output (a supervised class
probability or decision score, an anomaly score, or a reconstruction or
prediction error, depending on the learning paradigm). Let $Z_t$ collect the
recording and operating-context variables: recording membership, campaign,
operating condition and acquisition time. The intended detection path is
\begin{equation}
A_t \longrightarrow X_t \longrightarrow S_t ,
\end{equation}
However, the observed score--label association may also be consistent with
\begin{equation}
Z_t \longrightarrow X_t \longrightarrow S_t ,
\qquad Z_t \leftrightarrow Y_t ,
\end{equation}
because normal and attack behavior are often captured in separate recordings,
so labels co-vary with $Z_t$. These arrows represent explanatory alternatives,
not identified causal relationships. A strong association between $S_t$ and
$Y_t$ alone does not show that a detector responds specifically to attack
behavior: the same benchmark result can be consistent with attack-related
evidence, recording-associated differences, or both (Fig.~\ref{fig:protocol}a).
The four complementary steps of recording-aware evaluation are recording checks
and same-recording comparison, forward evaluation using only prior data,
temporal alignment tests, and operational alarm evaluation
(Fig.~\ref{fig:protocol}b).

Throughout, future observations are not used for preprocessing, model fitting,
calibration, or threshold selection. Decisions are made on half-open windows
$[w,w+\Delta)$, and an alert raised on a window is timestamped at its start $w$.

\subsection{Recording Checks and Same-Recording Evaluation}
\label{sec:session}
Recording-aware evaluation uses complementary comparisons to determine whether
detector performance is sensitive to differences between recordings. The
pre-attack recording check compares the benign recording with windows
from the attack recording that occur before the first labeled attack. Because
these windows contain no labeled attack activity, strong discrimination shows
that the recordings are distinguishable before the attack begins.

Same-recording evaluation compares attack windows with non-attack
windows from the same attack recording. To reduce contamination from activity
close to labeled actions, we apply a fixed temporal buffer and repeat the
comparison after excluding nearby background windows.

\begin{table}[t]
\centering
\caption{Recording checks and same-recording evaluation.}
\label{tab:diagnostics}
\footnotesize
\setlength{\tabcolsep}{3pt}
\begin{tabular}{@{}>{\raggedright\arraybackslash}p{0.27\columnwidth}>{\raggedright\arraybackslash}p{0.33\columnwidth}>{\raggedright\arraybackslash}p{0.32\columnwidth}@{}}
\toprule
\hd{Check} & \hd{What is compared} & \hd{What it tells us} \\
\midrule
Pre-attack recording check & benign recording vs.\ attack-recording windows before the first labeled attack & whether recordings are distinguishable before attack activity \\
Same-recording evaluation & attack windows vs.\ buffered non-attack windows from the same recording & attack-vs.-background discrimination without changing recordings \\
Cross-to-within gap $\Delta_{\mathrm{CW}}$ & cross-recording AUC minus same-recording AUC & sensitivity of apparent performance to recording structure \\
\bottomrule
\end{tabular}
\end{table}

To summarize how much apparent performance changes when the background
comparison changes from a separate recording to the same recording, we define
the cross-to-within gap as
\begin{equation}
\Delta_{\mathrm{CW}}=\mathrm{AUC}_{\mathrm{cross}}-\mathrm{AUC}_{\mathrm{within}}.
\end{equation}
A positive $\Delta_{\mathrm{CW}}$ indicates that discrimination is stronger
when attack windows are compared with a separate benign recording than when
they are compared with background from the same recording.
For each detector and evaluation step, the two comparisons use identical
scores and attack-positive windows; only the background population changes.
The gap is descriptive, not a causal estimate of a recording effect.
Same-recording background may contain unlabeled persistence, attack
consequences, or recovery, which can also reduce within-recording AUC.
Temporal buffers limit proximity to labeled actions but cannot exclude effects
lasting longer than the tested buffers.

\noindent\textbf{Recording-membership probe.} The probe is separate from IDS detection: it predicts the
two recording identities using only benign-reference and pre-attack windows
from the attack recording. It uses a time-ordered 70/30 train--test split within
each recording. Labels, timestamps, recording identifiers, availability
indicators, and raw host or network identifiers are excluded from the inputs.
Feature-family refits test whether separability is concentrated in a particular
representation.

\noindent\textbf{Operating-state check.}\label{sec:stateprobe}
Because differences in physical operating conditions could contribute to
recording separability, we also attempt to balance the recordings using the 22
physical-process variables. Pre-attack windows are matched to benign-reference
windows under a fixed caliper, and the comparison proceeds only if predefined
support, retention, pair-count, and covariate-balance requirements are
satisfied. The primary balance criterion is a maximum absolute standardized mean
difference of 0.10~\cite{austin2009balance}. If adequate balance cannot be
achieved, no matched detector-score comparison is performed.

\subsection{Forward and Temporal Evaluation}
\label{sec:forward}
Recording-aware evaluation also preserves the chronological structure of the
benchmark. Each test campaign is evaluated using only chronologically prior
data for preprocessing, model fitting, calibration, and threshold selection.
For ordered attack campaigns $C_1,\dots,C_K$, this produces a sequence of forward
evaluations, such as $C_1 \rightarrow C_2$ and $C_1{+}C_2 \rightarrow C_3$. In
each step, only earlier data construct the reference set and calibrate the
detector.

Reference provenance is part of this design. A reference can come from an
earlier benign recording or from action-screened history within the attack
recording; these choices are not interchangeable. In the latter case, labeled
attack windows are excluded, but unlabeled persistence or recovery may remain.
Section~\ref{sec:forwardsettings} distinguishes the references used in the
reported analyses. A detector fitted before the test period may remain fixed
across later campaigns; chronological evaluation does not itself require refitting.

\noindent\textbf{Temporal alignment check.}\label{sec:null}
Forward evaluation prevents future information from entering the detector, but
high scores may still reflect broad campaign conditions rather than the timing
of individual attack periods. We therefore keep detector scores fixed and
randomize the locations of labeled events within the held-out recording. The
randomized schedules preserve the number and duration of the observed events
while avoiding the true event intervals, surrounding buffers, and overlap
between pseudo-events. Circular shifts of the complete event schedule are also
used where sufficient background is available.

Let $T_{\mathrm{obs}}$ be the mean detector score within the true labeled
periods and let $T_j$ be the corresponding mean under randomized schedule $j$.
For $B$ randomized schedules, we report the randomization exceedance proportion
\begin{equation}
q_{\mathrm{align}} = \frac{1+\sum_{j=1}^{B}\mathbb{I}(T_j \geq T_{\mathrm{obs}})}{B+1}.
\end{equation}
Here, $\mathbb{I}(\cdot)$ is the indicator function. Smaller values indicate
stronger alignment of detector scores with the labeled
schedule. The test is interpreted as a temporal alignment check, not as evidence
that each individual attack action is identified separately.

\subsection{Operational Evaluation}
\label{sec:timingc}
Ranking metrics such as AUC describe how well detector scores separate labeled
attack and non-attack windows, but they do not directly describe the operational
behavior of the resulting alarms. We therefore evaluate event coverage and alarm
burden separately.

\noindent\textbf{Event detection.} The primary event-recall measure uses
chronological one-to-one matching between alarm episodes and labeled attack
events. Each alarm episode can receive credit for at most one overlapping attack
event, and each attack event can be matched to at most one alarm episode. This
prevents a single long alarm from being counted as multiple independent
detections simply because it overlaps several labeled actions. The rule is
deliberately stricter than range-aware time-series scoring that can assign
partial credit for overlap~\cite{tatbul2018}. For comparison,
we also report any-overlap recall, in which an attack event is considered
covered whenever it overlaps an alarm episode. This more permissive measure is
useful for describing coverage, but it is interpreted separately from primary
one-to-one event recall.
One-to-one recall is a deliberately conservative diagnostic for exposing credit
assigned to long alarm episodes; it is not proposed as a universal operational
definition for persistent campaigns.

\noindent\textbf{Alarm burden.} Alarm burden matters operationally in intrusion
detection~\cite{sok2023,axelsson2000}. We report outside-action occupancy,
defined as the fraction of non-action test windows that remain in alarm. Because
benchmark labels mark attack actions rather
than all possible consequences, persistence, or recovery, unmatched alarms are
not treated directly as conventional false positives.

\noindent\textbf{Evaluation units and fixed settings.}
\label{sec:opmetrics}
Attack campaigns are the independent generalization units, while repeated
random seeds are nested model fits rather than independent observations. We
therefore report results by forward step and seed range rather than treating all
campaign--seed combinations as independent samples. Models, preprocessing,
calibration procedures, thresholds, and primary evaluation criteria are fixed
before the corresponding held-out campaign is scored. Applying the procedure
requires sufficient reference support, multiple chronologically ordered
scenarios, temporal background for the controls, and event timing usable at
event level.

\noindent\textbf{Synthetic validation.}
\label{sec:controlvalidation}
The synthetic experiments confirm the intended behavior of the two diagnostics
under known score-generating mechanisms. One holds attack evidence fixed while
varying recording-associated variation; the other contrasts action-aligned,
campaign-constant, episode-envelope, and random score structures under the
temporal control of Sec.~\ref{sec:null}. In 1{,}000 repetitions of the first
control, increasing recording strength from $\lambda=0$ to $\lambda=2$ raises
mean cross-recording AUC from 0.760 to 0.983 while mean same-recording AUC
remains 0.760--0.761. In the temporal control, a campaign-constant score does
not exceed either randomized baseline in any forward step, whereas an
episode-envelope score exceeds both randomized baselines in all three steps.
At $C_3$, none of the 200 duration-matched or 200 circular-shift schedules
attains a statistic as large as the observed episode-envelope score. Thus, the
test distinguishes unrelated campaign membership from temporal structure, but
it is also sensitive to the broader attack-episode envelope. Full generator
definitions, parameters, and results are provided in the artifact.

\begin{table}[t]
\centering
\caption{Synthetic validation of the evaluation checks.}
\label{tab:synthetic}
\footnotesize
\setlength{\tabcolsep}{3pt}
\begin{tabular}{@{}>{\raggedright\arraybackslash}p{0.34\columnwidth}>{\raggedright\arraybackslash}p{0.30\columnwidth}>{\raggedright\arraybackslash}p{0.27\columnwidth}@{}}
\toprule
\hd{Control} & \hd{Expected behavior} & \hd{What it verifies} \\
\midrule
Add recording variation while holding attack evidence fixed & Cross-recording AUC rises; same-recording AUC is stable & $\Delta_{\mathrm{CW}}$ responds to recording-associated information \\
Change score timing relative to events & Randomization separates action-aligned from unrelated score timing & Schedule test measures alignment with labeled periods \\
\bottomrule
\end{tabular}
\end{table}
\FloatBarrier

\section{Experimental Setup}
\label{sec:case}

\subsection{ProvICS Dataset and Attack Campaigns}
ProvICS is a hardware-in-the-loop OT dataset that provides synchronized cyber
and physical observations from an industrial control environment
\cite{provics2026}. The testbed includes supervisory computing, PLC-based
control, Modbus/TCP communication, and a software-based chemical-reactor digital twin. Four
synchronized views are available on a common timeline: supervisory-host
provenance (M1), PLC provenance (M2), decoded Modbus traffic (M3), and
physical-process telemetry (M4). This multimodal structure enables us to
examine recording sensitivity across cyber and physical evidence and under
different fusion strategies.

Table~\ref{tab:dataset} summarizes the synchronized views and campaigns used
in this study.

\begin{table}[H]
\centering
\caption{ProvICS modalities and attack campaigns~\cite{provics2026,provicsdata2026}.}
\label{tab:dataset}
\footnotesize
\setlength{\tabcolsep}{2.8pt}
\begin{tabular}{@{}l>{\raggedright\arraybackslash}p{0.27\columnwidth}>{\raggedright\arraybackslash}p{0.235\columnwidth}>{\raggedright\arraybackslash}p{0.26\columnwidth}@{}}
\toprule
\multicolumn{4}{@{}l}{Panel A: synchronized views}\\
\hd{View} & \hd{Evidence source} & \hd{Source format} & \hd{Availability (\%)}\\
\midrule
M1 & supervisory-host provenance & event/provenance graph & 94.9\%\\
M2 & controller/PLC provenance & event/provenance graph & 77.6\%\\
M3 & Modbus protocol semantics & protocol-event sequence$^{\ast}$ & 94.9\%\\
M4 & physical process & 22-variable time series, 1\,Hz & 98.7\%\\
\midrule
\multicolumn{4}{@{}l}{Panel B: attack campaigns}\\
\hd{Campaign} & \hd{Events / positive windows (60\,s)} & \hd{Campaign windows} & \hd{Campaign profile}\\
\midrule
$C_1$ & 6 / 12 & 409 & aggressive initial intrusion\\
$C_2$ & 9 / 47 & 288 & low-and-slow\\
$C_3$ & 6 / 21 & 254 & controller sabotage\\
$C_4$ & 11 / 48 & 295 & multi-stage persistent campaign\\
\bottomrule
\end{tabular}

{\footnotesize $^{\ast}$The parsed release contains no register-value field, so
M3 is a reduced-input protocol-semantic view. Availability is the fraction of
the aligned 60-s master grid for which the view is present.}
\end{table}

The analyzed release contains separate benign and attack recordings. At 60-s
resolution, the benign recording contains 51.95\,h of data (3{,}117 windows),
while the attack recording contains 20.77\,h (1{,}246 windows). Because both
attack and non-attack periods are available within the attack recording,
ProvICS supports conventional cross-recording evaluation and the same-recording
comparison required by our procedure.

\label{sec:campaigns}
The attack recording contains 32 labeled actions organized into four
chronologically ordered campaigns: an aggressive initial intrusion ($C_1$),
low-and-slow activity ($C_2$), controller sabotage ($C_3$), and a multi-stage
persistent campaign ($C_4$)~\cite{provics2026,mitreics2026}. These campaigns
provide the temporal ordering required for forward evaluation. ProvICS labels
identify attack actions rather than every possible consequence, persistence
period, or recovery state. We therefore label a window positive when it overlaps
a labeled attack interval, but do not automatically treat every other alarm as a
conventional false positive. Across $C_2$--$C_4$, evaluation covers 26 unique
attack events. Repeated model fits are treated as nested runs rather than
independent observations.

\subsection{Detectors and Fusion Methods}
\label{sec:detectors}
Each modality is divided into non-overlapping windows of
$\Delta\in\{5,10,30,60\}$ seconds. A window is labeled as an attack window
when it overlaps a labeled attack interval. The same feature families and
labeling rule are used at all resolutions, so window-size effects can be
examined without changing the underlying representation.

For the cyber modalities, window-level features summarize activity rather than
use raw identifiers. M1 and M2 contain provenance-based counts, rates, graph
statistics, timing information, and semantic event categories. M3 summarizes
Modbus activity through function-code and register-reference statistics,
read/write behavior, and timing features. M4 contains 199 features: nine
within-window summaries for each of the 22 physical-process variables, plus the
sample count. Raw IP addresses, ports,
transaction identifiers, process IDs, hashes, command lines, paths, timestamps,
labels, and availability indicators are excluded from model inputs.

The primary analysis intentionally uses a simple fixed Isolation Forest for each
modality. The objective is not to identify the best-performing IDS, but to
generate comparable detector scores for evaluating recording sensitivity under a
common procedure. Each Isolation Forest uses 300 trees with a maximum sample
size of 256 and is evaluated using five fixed random seeds. Campaign-wise
summaries treat these as nested repeated fits. For modality $m$, $s_m(t)$ is
the negative Isolation Forest \texttt{score\_samples} output, so larger values
indicate greater anomaly. Scores are standardized from the prior reference data:
\[
z_m(t)=\frac{s_m(t)-\mu_m}{\sigma_m}.
\]

We evaluate the four individual modalities together with score-level fusion
methods. Cyber-only fusion sums standardized scores for M1--M3, while all-view
fusion sums scores for M1--M4. Late-OR fusion uses the maximum standardized
score across the four modalities. We also evaluate a one-component canonical
correlation analysis (CCA) residual that models the calibration-time
relationship between the cyber-score vector and the physical score; the fused
anomaly score is the absolute discrepancy between their canonical variates.
None of the primary detectors or fusion methods uses attack labels as a
training target.
Attack labels define evaluation windows and screen attack-recording history
when that reference is used. Per-view AUCs use modality-specific availability
masks, not a shared complete-case set. Fusion uses aligned score streams:
missing scalar features receive training medians and absent categorical counts
are zero before scoring. Sum-$z$ and late-OR then combine these scores without
a joint availability mask; CCA is fitted on finite calibration scores.

\subsection{Forward Evaluation Settings}
\label{sec:forwardsettings}
The primary evaluation uses 5, 10, 30, and 60-s windows with a fixed 300-s
exclusion buffer around labeled attack intervals. The 60-s extraction is
verified against the persisted canonical ProvICS records; detailed reconstruction
and merge procedures are provided in the artifact.

The window-size and fusion comparisons fit preprocessing and Isolation Forest
on the first 70\% of the separate benign recording. The remaining 30\% supplies
cross-recording evaluation negatives; attack positives and buffered background
come from the attack recording. Standardization and thresholds use benign
training-partition scores. For these comparisons, the calibration reference
reuses the benign fitting partition rather than a disjoint calibration set.
These fixed benign-reference fits are also scored
by campaign; they are not refitted on earlier attack-recording background.

The separate audit of action-screened attack-recording history gives
fit/calibration counts of 278/120 for $C_2$, 358/154 for $C_3$, and 396/170 for
$C_4$ at 60~s. These describe an alternative reference construction, not the
benign-reference partitions behind Figs.~\ref{fig:confound}--\ref{fig:frontier}.
Forward campaign checks evaluate $C_2$--$C_4$; $C_1$ supplies the first past
attack examples for the supervised check, rather than a held-out campaign.

The primary alarm threshold is the empirical 98.6th percentile of calibration
scores, corresponding to a nominal 1.4\% calibration exceedance rate. This
matches the false-positive rate reported for the ProvICS baseline
\cite{provics2026}; we retain it as a fixed reference operating point set
before held-out scoring rather than selecting it from held-out performance. We
also evaluate a fixed operating range from 0.5\% to 10\% exceedance to characterize
alarm behavior across thresholds without selecting a threshold after observing
test performance. AUC is a threshold-free ranking measure, whereas thresholded
results characterize operational alarm behavior. At 60~s, 1{,}000 moving-block
bootstrap replicates with 10-window blocks resample the benign fitting/calibration
partition, rederive the threshold, and apply it to unchanged held-out scores to
assess calibration sensitivity. Temporal alignment checks use 2,000
duration-matched randomized schedules and circular schedule shifts where
sufficient background is available.

\subsection{Supervised Model Check}
\label{sec:robustsetup}
To determine whether observed recording-related differences depend on the primary Isolation
Forest detector, we repeat the cross-recording and same-recording comparisons
with three fixed supervised models: logistic regression, random
forest, and a small multilayer perceptron. Training negatives come from the
benign training partition; positives are labeled actions from earlier campaigns
only. The held-out campaign contributes no training examples. No
hyperparameter search uses the held-out campaign. Random forest and
multilayer-perceptron fits use five fixed random seeds nested within each campaign.

For each fitted model, cross-recording and same-recording AUCs use the same
detector scores and attack windows; only the background population changes. This
measures sensitivity to background selection while keeping the fitted model and
positive samples fixed. The input features follow the same fixed
screening rules as the primary analysis and exclude identifier-derived
information.

\subsection{External Check on HAI}
\label{sec:externalsetup}
We use HAI~21.03~\cite{hai2021} as an external check of whether the recording
sensitivity observed on ProvICS also appears in another OT benchmark. For each
eligible comparison, an Isolation Forest is fitted using only the
chronologically prior portion of a normal reference file and applied to a
compatible attack file. Cross-recording and same-recording AUCs use the same
fitted detector and attack-positive windows; only the background population
changes.

Three normal reference files and five attack files provide candidate
reference--target combinations. A comparison is included only when sufficient
attack and buffered non-attack windows are available and the data provide usable
chronology and event timing: at least 20 attack-positive and 20 buffered
non-action windows. The HAI analysis is a qualitative external check rather
than a direct performance comparison with ProvICS. Because the datasets use
different system configurations, representations, and temporal structures, we
compare the presence and direction of recording-related changes rather than
absolute detector performance.

\subsection{Reproducibility}
\label{sec:repro}
All dataset versions, feature-screening rules, model configurations,
calibration settings, thresholds, and primary evaluation criteria are fixed
before the corresponding held-out campaign is scored. The configurations are
recorded and SHA-256 hashed so that later changes can be documented separately.
The accompanying artifact provides source code, detector scores, window
metadata, random seeds, feature definitions, temporal-randomization procedures,
checksummed result tables, an execution registry, and verification of the 60-s
ProvICS extraction against persisted canonical records.

\section{Results}
\label{sec:results}

The synthetic experiments confirm the intended behavior of the two diagnostics
under known score-generating mechanisms: recording information can increase
cross-recording AUC while leaving same-recording discrimination unchanged, and
schedule randomization detects temporal alignment but can also respond to
episode-wide elevation without action-specific detection.
Generator details and all synthetic results are in the artifact.

\subsection{Recording-Associated Differences Across Modalities}
\label{sec:confound}

\begin{figure*}[t]
\centering
\includegraphics[width=\textwidth]{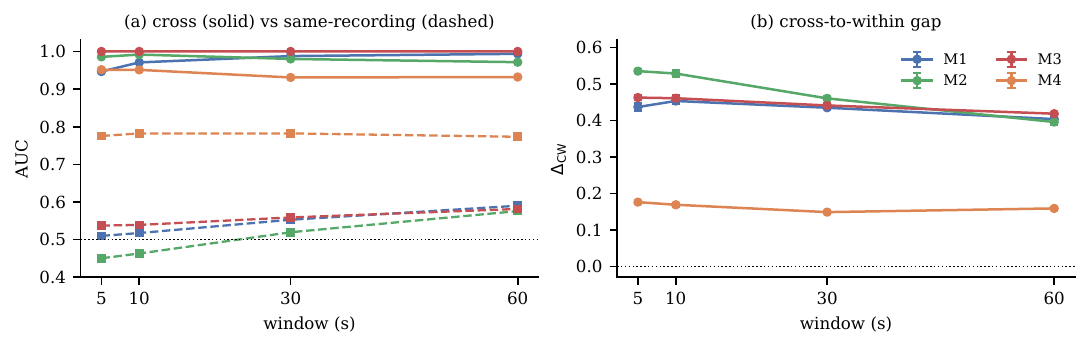}
\caption{Window-size sensitivity of the per-view comparison. (a)
Cross-recording (solid) and same-recording (dashed) AUC. (b) The
cross-to-within gap $\Delta_{\mathrm{CW}}$. The four window sizes are
correlated re-aggregations of one recording pair, not independent replications.}
\label{fig:confound}
\end{figure*}

The pre-attack recording check shows that the cyber modalities
distinguish the recordings before any labeled attack occurs. It contains only
benign-reference windows and attack-recording windows that occur before the
first labeled attack. It uses the benign-trained Isolation Forest anomaly score
to distinguish these pre-attack windows from held-out benign windows, not the
separate membership classifier. Across 5, 10, 30, and 60-s windows, the pre-attack recording-check AUC is
0.944--0.998 for the three cyber views and 0.763--0.786 for M4.
A separate recording-membership probe, trained on
the earlier 70\% of each recording and tested on the later 30\%, reaches AUC
1.000 for M1--M3 with both logistic regression and random forest at every
resolution. These results show that the cyber representations contain strong
information about recording identity before the first labeled attack begins.

This ordering changes under same-recording evaluation. The
cross-to-within gap is positive for every modality at every studied window
size, ranging from 0.149 to 0.535. Under cross-recording
evaluation, M3 is the strongest individual modality at all four resolutions.
When attack windows are compared instead with background from the same
recording, M4 becomes the strongest modality and has the smallest
recording-related change (0.149--0.176). The ranking reversal persists across
all studied window sizes; these resolutions are sensitivity checks based on
re-aggregations of the same recordings, not independent replications.
Figure~\ref{fig:confound} shows the cross-recording and same-recording AUCs
together with $\Delta_{\mathrm{CW}}$.

The result is stable under wider action buffers. Holding the detector,
attack-positive windows, and cross-recording background fixed, we repeated the
same-recording comparison with 300-, 600-, and 900-s event-based exclusions.
All 16 modality--window settings retain a positive seed-median
$\Delta_{\mathrm{CW}}$ at every buffer (Appendix~\ref{app:buffer}). Moving from
300 to 900~s removes 6--9\% of the same-recording background and changes the
median gap by less than 0.009. M4 remains the least recording-sensitive view at
every resolution. The buffer analysis therefore supports the stability of the
gap over the tested buffers; longer attack consequences and the source of the
remaining difference remain unresolved.

Operating-state differences remain a plausible contributor.
Recording separability is distributed across several cyber feature families;
removing any one changes held-out recording-classification performance only
marginally. We also attempted to balance the two recordings using the 22
physical-process variables, but the pre-specified balance criterion was not
satisfied at any resolution. The maximum post-match absolute standardized mean
difference remained between 0.25 and 0.47, above the 0.10 threshold, and
7--21 of the 22 variables remained unbalanced depending on window size. A
physical-state-only recording probe achieved AUC 0.834--0.875. These
observations indicate that operating-state differences may contribute to the
recording-associated differences, but the available data do not allow them to
be attributed to a single source.

\subsection{Fusion and Operational Performance}
\label{sec:frontier}

\begin{figure*}[t]
\centering
\includegraphics[width=\textwidth]{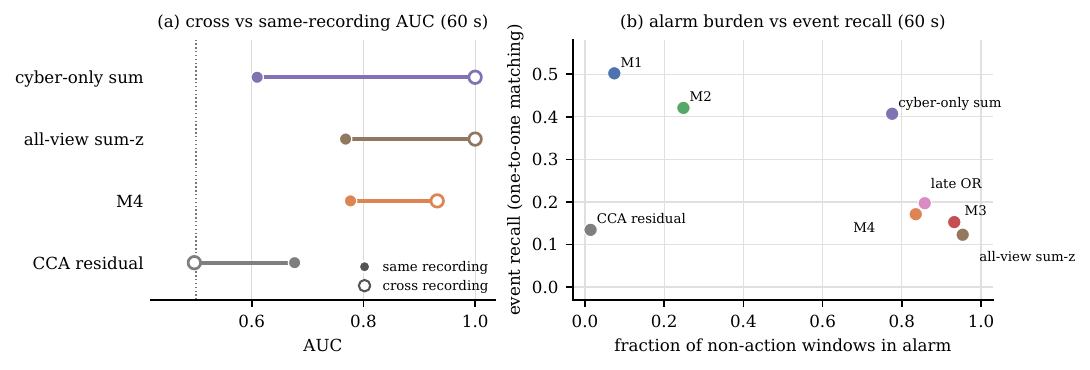}
\caption{Fusion and operational performance at 60~s. (a) Cross-recording
(open) and same-recording (filled) AUC for selected score families; connecting
segments encode the cross-to-within gap. (b) One-to-one event recall
versus the fraction of non-action windows in alarm at the nominal 1.4\%
calibration budget.}
\label{fig:frontier}
\end{figure*}

Cyber-only fusion has the largest cross-to-within gap at 5, 10, and
30~s, with values of 0.540, 0.528, and 0.453, respectively. At 60~s, the gap remains
substantial at 0.390. Cyber-only fusion can therefore amplify sensitivity to
recording-associated differences shared across the cyber views.

Adding the physical modality changes this behavior. All-view sum-$z$ reduces
the gap to approximately 0.23 and improves same-recording AUC at every studied
resolution. However, this ranking improvement comes with substantial alarm
burden: at the nominal calibration setting, the detector remains in alarm
during 93--95\% of non-action windows. Other fusion strategies behave
differently. For example, the CCA residual has smaller or negative
recording-related gaps at 30 and 60~s, but its event recall remains low
(Fig.~\ref{fig:frontier}). Fusion should therefore be evaluated not only by
AUC, but also by its sensitivity to recording structure and its operational
alarm behavior. The nominal rate describes the benign calibration scores, not
a guaranteed test alarm rate. This calibration-to-test mismatch does not, by
itself, establish progressive drift within the attack recording.

Long alarms can inflate apparent event coverage. At 60~s,
all-view sum-$z$ achieves any-overlap event recall of 0.963, but its stricter
one-to-one event recall is only 0.123. Across the reported forward steps and
nested fits, it produces a mean of 1.3 alarm episodes, and the longest spans
227 consecutive 60-s windows (3.8\,h). Under the permissive
any-overlap rule, this single long episode receives credit for multiple labeled
actions. A detector that remains in alarm across a long period may overlap many
attacks without repeatedly identifying their individual occurrence. We
therefore interpret one-to-one event recall together with the fraction of
non-action time spent in alarm as the primary operational view of thresholded
results.

\subsection{Model, Temporal, and External Checks}
\label{sec:supporting}
\label{sec:ablations}

The supervised-model check retains positive cross-to-within AUC gaps
for M1--M3 across the studied model families. This pattern is
not unique to the primary Isolation Forest. M4 remains more
dependent on model family and attack campaign.

Temporal alignment and calibration choice also affect the
interpretation of detector performance. In the temporal randomization test,
M4 and all-view sum-$z$ show stronger alignment with the true labeled attack
periods than M2, M3, and cyber-only fusion. At 60~s, the duration-matched
randomization exceedance proportion for M4 and all-view sum-$z$ attains its
minimum possible value, $q_{\mathrm{align}}=1/2001\approx0.0005$, for all five
seeds, compared with mean values of 0.1278 for M3 and 0.1865 for cyber-only
fusion. Because
operating-state matching
did not satisfy the pre-specified balance criterion, we do not construct a
regime-aware randomized baseline. At 60~s, rederiving the threshold across
moving-block bootstrap samples of the calibration partition produces
outside-action occupancy of 0.13--0.94 for cyber-only fusion and 0.79--0.98 for
all-view sum-$z$ at the nominal 1.4\% calibration exceedance rate. Thus,
calibration-sample variability can materially affect operational behavior even
when the fitted model and held-out scores are unchanged.

Table~\ref{tab:hai} reports all 12 eligible HAI reference--target
comparisons. Its cross-to-within differences are substantially smaller than on
ProvICS, and the sign changes with the selected normal reference for two of the
four eligible attack files. HAI therefore establishes neither a universal
direction nor a universal magnitude of recording sensitivity. Instead, the
external check shows that the quantity is dataset- and reference-dependent and
should be measured explicitly rather than assumed.
\label{sec:externalresults}

\section{Discussion}
\label{sec:discussion}

\subsection{Interpreting Benchmark Performance}
High benchmark performance shows that a detector separates the labeled classes
under a particular evaluation design. It does not, by itself, show which
characteristics of the data produced that separation. This distinction becomes
important when normal and attack samples are drawn from different recordings.

On ProvICS, the interpretation changes substantially when the background
comparison is moved from a separate benign recording to the same attack
recording. Under cross-recording evaluation, the cyber modalities appear
strongest, while same-recording evaluation reverses the ranking and makes the
physical modality the strongest individual view. The strong pre-attack
separability further shows that apparent discrimination is sensitive to
recording structure, not only to labeled attack periods.

The two evaluations therefore answer different questions. Cross-recording
evaluation asks whether the detector separates the benchmark classes as they
are provided. Same-recording evaluation asks whether attack windows remain
distinguishable from background when the recording boundary is removed from the
comparison. Reporting both provides a more informative interpretation of
detector performance.

Recording-aware evaluation adds background-source sensitivity to temporal,
unseen-attack, and cross-network checks~\cite{arp2022,tesseract2019,kus2022}.
The supervised results show that the cyber-side gaps are not unique to
Isolation Forest, while the physical-view results depend on model and campaign.

\subsection{Implications for Multimodal OT IDS}
Multimodal fusion can amplify sensitivity to recording-associated differences
when several modalities share them, so improved aggregate performance does not
necessarily reflect stronger attack-related evidence. On ProvICS, cyber-only fusion has a
larger cross-to-within gap than any individual cyber modality at three of the
four studied window sizes. Adding the physical modality reduces that gap and
improves same-recording discrimination, but the all-view fusion remains in
alarm during most non-action windows at the nominal operating point.

Fusion should therefore be assessed under both cross-recording and
same-recording settings, with ranking metrics reported alongside event recall
and alarm burden. More broadly, a modality's contribution depends on the
evaluation setting. In ProvICS, the physical view appears weaker under
cross-recording evaluation but becomes the strongest individual view under
same-recording evaluation.

\subsection{Implications for Benchmark Design}
The results also have implications for the design and use of OT
intrusion-detection benchmarks. When normal and attack data are collected in
separate recordings, recording identity should be preserved as metadata so that
potential recording-related differences can be examined during evaluation. At
the same time, raw identifiers or other features that directly reveal recording
membership should be excluded from detector inputs unless they are explicitly
part of the intended detection task.

Where possible, benchmark evaluations should report both conventional
cross-recording performance and a same-recording comparison between attack and
background periods. This allows researchers to determine whether strong
performance remains when the recording boundary is removed from the comparison.
Chronological splits should also be preserved so that preprocessing, model
fitting, calibration, and threshold selection rely only on data available before
the held-out test period.

Benchmark design can further support stronger evaluation by providing multiple
benign and attack recordings, repeated operating conditions, and sufficient
non-attack background within recordings that contain attacks. Embedding attacks
within longer periods of continuous normal operation is particularly useful
because it enables attack and background behavior to be compared under the same
recording conditions.

The contrast between ProvICS and HAI illustrates why these checks should be
performed rather than assumed. ProvICS shows a large recording-related change,
whereas HAI shows smaller and reference-dependent recording sensitivity. Recording
sensitivity may therefore vary across datasets, and benchmark structure should
be treated as part of the evaluation rather than as a neutral property of the
data.

\subsection{Limitations and Generalizability}
\label{sec:threats}
This study has several limitations. First, the main empirical analysis is based
on one multimodal testbed, ProvICS, with four attack campaigns and three
chronologically ordered forward evaluation steps. The results therefore characterize this
dataset and evaluation setting rather than the prevalence of recording-related
differences across OT benchmarks. Window sizes and random seeds are not
independent replications. Campaign composition, including process-targeting
actions, may influence the aggregate modality ranking; campaign-level
uncertainty is not established by the reported seed variation.

Second, the observed recording differences cannot be attributed to a single
source. Differences in operating regime, acquisition conditions, system
configuration, or security-relevant activity outside the labeled attack
intervals remain plausible contributors. Our attempt to balance the recordings
using the 22 physical-process variables did not satisfy the pre-specified
covariate-balance requirement at any studied resolution. The analysis therefore
shows that measured performance is sensitive to recording structure, but it does not
determine the mechanism responsible for that difference. Feature-family
ablations do not identify a unique recording artifact or rule out cumulative
or uptime-related proxies.

Third, several measurement and evaluation constraints remain. Alarm behavior is
sensitive to calibration-sample variability, and the available calibration
partitions provide limited support for estimating extreme thresholds. In addition, the
supervised check covers one linear, one tree-based, and one neural model family
rather than the full range of possible OT IDS architectures. Unequal modality
availability and M3's missing register values further limit the cyber--physical
comparison to this released representation. Common-window comparisons,
pre-attack-only background controls, and campaign-level uncertainty estimates
are left for future work.

Finally, the external HAI analysis uses a dataset-specific representation and
temporal resolution and is intended as a qualitative comparison rather than a
direct cross-dataset performance benchmark. Recording-aware evaluation also does
not address all deployment challenges, including operating-envelope coverage,
concept or operating-regime drift, and discrimination between attacks and
process faults. These remain important directions for broader validation.

\section{Conclusion}
\label{sec:conclusion}

This study examined how recording structure can change the interpretation of
multimodal OT intrusion detection. On ProvICS, the cyber modalities distinguish
the benign and attack recordings before the first labeled attack, and
same-recording evaluation reverses the apparent modality ranking, with the
physical view becoming the strongest individual modality. Cyber-only fusion can
further amplify recording-related differences, while operational evaluation
shows that high apparent event coverage can be driven by long alarm episodes and
substantial alarm activity outside labeled attack periods.

These findings motivate a more careful interpretation of benchmark performance.
Reporting only cross-recording AUC or similar classification metrics may give an
incomplete picture when normal and attack data come from different recordings.
Recording-aware evaluation adds complementary same-recording, forward,
temporal, and event-level checks that help determine whether reported
performance remains meaningful under a more controlled comparison.

The model-agnostic procedure measures sensitivity to background selection
alongside temporal alignment and alarm behavior; it does not identify a causal
recording effect.

More broadly, OT IDS benchmarks should preserve recording information for
evaluation, include sufficient within-recording background, and, where possible,
embed attacks within continuous operation across repeated recordings. These
practices can make benchmark results more informative and help assess whether
reported detection performance depends strongly on recording structure.

\appendices
\section{External HAI Reference--Target Comparisons}
\label{app:hai}

Table~\ref{tab:hai} gives every HAI comparison meeting the pre-specified
minimum of 20 attack-positive and 20 buffered same-recording background
windows. Each row uses one fixed Isolation Forest and the same attack-positive
windows for the cross-recording and same-recording AUCs; only the background
population changes. The twelve rows are dependent repeated settings because
they reuse three reference files and four target files. Test1 is excluded for
insufficient attack support (14 positive windows).

\begin{table}[H]
\centering
\caption{HAI external check. $\Delta_{\mathrm{CW}}=$ cross-recording AUC minus
same-recording AUC. Rows sharing a target or reference are not independent
replications.}
\label{tab:hai}
\footnotesize
\setlength{\tabcolsep}{4.2pt}
\begin{tabular}{@{}llrrr@{}}
\toprule
\hd{Target} & \hd{Reference} & \hd{Cross AUC} & \hd{Same AUC} & \hd{$\Delta_{\mathrm{CW}}$} \\
\midrule
test2 & train1 & 0.815 & 0.879 & -0.064 \\
test3 & train1 & 0.979 & 0.957 & \phantom{-}0.022 \\
test4 & train1 & 0.945 & 0.926 & \phantom{-}0.019 \\
test5 & train1 & 0.963 & 0.921 & \phantom{-}0.042 \\
test2 & train2 & 0.765 & 0.797 & -0.032 \\
test3 & train2 & 0.964 & 0.898 & \phantom{-}0.066 \\
test4 & train2 & 0.855 & 0.882 & -0.027 \\
test5 & train2 & 0.906 & 0.876 & \phantom{-}0.030 \\
test2 & train3 & 0.896 & 0.862 & \phantom{-}0.034 \\
test3 & train3 & 0.991 & 0.979 & \phantom{-}0.012 \\
test4 & train3 & 0.979 & 0.952 & \phantom{-}0.026 \\
test5 & train3 & 0.980 & 0.945 & \phantom{-}0.035 \\
\bottomrule
\end{tabular}
\end{table}
\FloatBarrier

\section{Background-Buffer Sensitivity}
\label{app:buffer}

\begin{table*}[t]
\centering
\caption{Same-recording buffer sensitivity. Entries are seed-median
$\Delta_{\mathrm{CW}}$ values. $n_{\mathrm{within}}$ gives the corresponding
seed-median number of retained same-recording background windows at 300/900
seconds. Detector, features, attack positives, and cross-recording negatives
are fixed throughout.}
\label{tab:buffer}
\footnotesize
\setlength{\tabcolsep}{4.5pt}
\begin{tabular}{@{}rrrrrr@{}}
\toprule
\hd{Window (s)} & \hd{View} & \hd{$\Delta_{\mathrm{CW}}$, 300 s} & \hd{$\Delta_{\mathrm{CW}}$, 600 s} & \hd{$\Delta_{\mathrm{CW}}$, 900 s} & \hd{$n_{\mathrm{within}}$ (300/900 s)} \\
\midrule
5  & M1 & 0.439 & 0.440 & 0.441 & 12{,}067/11{,}107 \\
5  & M2 & 0.538 & 0.544 & 0.546 & 9{,}895/9{,}214 \\
5  & M3 & 0.463 & 0.467 & 0.468 & 12{,}093/11{,}133 \\
5  & M4 & 0.174 & 0.174 & 0.174 & 12{,}629/11{,}676 \\
10 & M1 & 0.450 & 0.450 & 0.450 & 6{,}035/5{,}555 \\
10 & M2 & 0.535 & 0.541 & 0.544 & 4{,}949/4{,}607 \\
10 & M3 & 0.468 & 0.473 & 0.473 & 6{,}047/5{,}567 \\
10 & M4 & 0.167 & 0.167 & 0.166 & 6{,}315/5{,}838 \\
30 & M1 & 0.431 & 0.432 & 0.432 & 2{,}011/1{,}851 \\
30 & M2 & 0.459 & 0.465 & 0.468 & 1{,}653/1{,}537 \\
30 & M3 & 0.446 & 0.452 & 0.453 & 2{,}013/1{,}853 \\
30 & M4 & 0.147 & 0.147 & 0.147 & 2{,}103/1{,}944 \\
60 & M1 & 0.411 & 0.413 & 0.413 & 1{,}004/924 \\
60 & M2 & 0.395 & 0.399 & 0.402 & 829/770 \\
60 & M3 & 0.417 & 0.422 & 0.422 & 1{,}004/924 \\
60 & M4 & 0.159 & 0.160 & 0.159 & 1{,}051/971 \\
\bottomrule
\end{tabular}
\end{table*}

The buffer-sensitivity analysis changes only the event-based exclusion around
labeled actions. Detector configuration, five seeds,
attack-positive windows, and cross-recording benign background are fixed.
Table~\ref{tab:buffer} reports all settings; none is removed for lower support.
The 300-s replay reproduces the baseline to numerical precision. This analysis
evaluates sensitivity to the definition of the same-recording background;
attribution of the remaining score differences is outside its scope.

\FloatBarrier

\bibliographystyle{IEEEtran}
\bibliography{references}

\section*{LLM Usage Statement}
LLMs were used for editorial purposes in this manuscript, and all outputs were
inspected by the authors to ensure accuracy and originality.

\end{document}